\documentclass[journal]{IEEEtran}
\usepackage{amsmath,amsfonts}
\usepackage{algorithmic}
\usepackage{array}
\usepackage[caption=false,font=normalsize,labelfont=sf,textfont=sf]{subfig}
\usepackage{stfloats}
\usepackage{url}

\usepackage{verbatim}
\usepackage{graphicx}
\usepackage{pifont}
\usepackage{multirow}
\usepackage{cite}
\newcommand{\subhead}[1]{\vspace {1pt}\noindent{\textbf{#1.}}}
\def\BibTeX{{\rm B\kern-.05em{\sc i\kern-.025em b}\kern-.08em
    T\kern-.1667em\lower.7ex\hbox{E}\kern-.125emX}}
\usepackage{balance}
\usepackage{multicol}

\usepackage{tikz}
\usetikzlibrary{patterns}

\newcommand*\redcirc[1][1ex]{\tikz\fill (0,0) circle (#1);}

\newcommand*\fullcirc[1][1ex]{\tikz\draw[pattern=north east lines, pattern color=black] (0,0) circle (#1);}

\newcommand*\emptycirc[1][1ex]{\tikz\draw (0,0) circle (#1);}

\newcommand*\purplecirc[1][1ex]{\tikz\draw[lightgray,fill=lightgray] (0,0) circle (#1);}

\usepackage{pdfrender}

\usepackage{pifont}%

\usepackage{fontawesome5}

\usepackage{listings} %
\usepackage{color} %
\definecolor{mygreen}{rgb}{0,0.6,0}
\definecolor{mygray}{rgb}{0.5,0.5,0.5}
\definecolor{mymauve}{rgb}{0.58,0,0.82}
 
\definecolor{darkgray}{rgb}{.4,.4,.4}
\definecolor{purple}{rgb}{0.65, 0.12, 0.82}
 
\lstdefinelanguage{JavaScript}{
keywords={typeof, new, true, false, catch, function, return, null, catch, switch, var, if, in, while, do, else, case, break},
keywordstyle=\color{blue}\bfseries,
ndkeywords={class, export, boolean, throw, implements, import, this},
ndkeywordstyle=\color{darkgray}\bfseries,
identifierstyle=\color{black},
sensitive=false,
comment=[l]{//},
morecomment=[s]{/*}{*/},
commentstyle=\color{purple}\ttfamily,
stringstyle=\color{red}\ttfamily,
morestring=[b]',
morestring=[b]"
}
 
\usepackage{xparse}
\NewDocumentCommand\mm{g}{%
  \IfNoValueF{#1}{{\color{blue} \textbf{(MM: #1)}}}%
  \IfNoValueT{#1}{{\color{blue} \textbf{(MM)}}}%
}
\NewDocumentCommand\am{g}{%
  \IfNoValueF{#1}{{\color{red} \textbf{(AM: #1)}}}%
  \IfNoValueT{#1}{{\color{red} \textbf{(AM)}}}%
}

\NewDocumentCommand\BT{g}{%
  \IfNoValueF{#1}{{\color{purple} \textbf{(BT: #1)}}}%
  \IfNoValueT{#1}{{\color{purple} \textbf{(BT)}}}%
}

\begin{document}
\title{Security Weaknesses in IoT Management Platforms}
\author{Bhaskar~Tejaswi,~\IEEEmembership{Student Member,~IEEE,}
        Mohammad~Mannan,  and~Amr~Youssef,~\IEEEmembership{Senior~Member,~IEEE}
        
        \thanks{Bhaskar Tejaswi, Mohammad Mannan and Amr Youssef are with the Concordia Institute for Information Systems Engineering, Concordia University, Montreal, QC H3G 1M8, Canada (e-mail:
\{b\_tejasw,mmannan,youssef\}@ciise.concordia.ca).}
\thanks{This paper is an extended version of an ACM CODASPY 2023 paper~\cite{btcodaspy}.
This work is partially supported by the Natural Sciences and Engineering Research Council of Canada (NSERC).}}

\markboth{IEEE INTERNET OF THINGS JOURNAL ,~Vol.~00, No.~0, ABC~2023}
{Shell \MakeLowercase{\textit{et al.}}: Bare Demo of IEEEtran.cls for IEEE Journals}

\maketitle

\begin{abstract}
A diverse set of Internet of Things (IoT) devices are becoming an integrated part of daily lives, and playing an increasingly vital role in various industry, enterprise and agricultural settings. The current IoT ecosystem relies on several IoT management platforms to manage and operate a large number of IoT devices, their data, and their connectivity. Considering their key role, these platforms must be properly secured against cyber attacks. In this work, we first explore the core operations/features of leading platforms to design a framework to perform a systematic security evaluation of these platforms. Subsequently, we use our framework to analyze a representative set of 52 IoT management platforms, including 42 web-hosted and 10 locally-deployable  platforms. We discover a number of high severity unauthorized access vulnerabilities in 9/52 evaluated IoT management platforms, which could be abused to perform attacks such as remote IoT SIM deactivation, IoT SIM overcharging and IoT device data forgery. More seriously, we also uncover instances of broken authentication in 13/52 platforms, including complete account takeover on 8/52 platforms along with remote code execution on 2/52 platforms. In effect, 17/52 platforms were affected by vulnerabilities that could lead to platform-wide attacks.
\textcolor{black}{Overall, vulnerabilities were uncovered in 33 platforms, out of which 28 platforms responded to our responsible disclosure.} We were also assigned 11 CVEs and awarded bounty 
for our findings.
\end{abstract}


\section{Introduction}
\IEEEPARstart{T}{he} pace of IoT adoption is rapidly increasing. It is estimated that the number of IoT  devices will reach 38.6 billion worldwide by 2025 and 50 billion by 2030~\cite{pasteward_2021}. IoT devices play a significant role in our daily lives (e.g., home automation), as well as at the enterprise level (e.g., device fleet management).
A key component of the IoT ecosystem is an IoT platform, which hosts a number of endpoints supporting the business operations utilizing IoT devices.
Some of these platforms provide data management services to enable  data collection from the IoT devices, followed by its processing and analytics. Some platforms offer device management services for users (e.g., enterprise IoT device administrators)
to remotely connect to their devices by using a platform's web portal and APIs. %
In others, users are also allowed to remotely execute commands on the IoT devices (by installing an edge client on the devices), and to upload firmware files to IoT devices. Such versatile functionalities, if not properly designed/implemented, can result in serious security issues. In 2021, researchers at FireEye disclosed a critical vulnerability in the device onboarding process of the Kalay cloud platform~\cite{fisher2021}, allowing a remote attacker to collect the login credentials and execute commands on millions of IoT devices managed by the platform. 
Beyond WiFi/LAN, IoT devices are increasingly being internet-connected via cellular networks.
Cellular IoT SIM sellers provide connectivity management platforms~\cite{emnifydefinition}, to facilitate their customers to easily and efficiently manage all of their IoT SIM cards 
(e.g., remotely activate/deactivate SIM cards).
IoT connectivity services can also be abused to compromise IoT devices, and to commit financial frauds and criminal activities (see e.g.,~\cite{iot-conn3, iot-conn2, iot-conn1}).

The term ``IoT platform'' is used by different vendors offering various combinations of services. Given the heterogeneous nature of platforms, to perform a systematic evaluation, we focus on three key IoT management services: connectivity, device, and data management. We define an IoT management platform as a platform that provides either one or a combination of these services for IoT devices. These platforms can be used by enterprises for the devices used by them, or businesses that sell IoT devices to consumers. Such platforms can be \emph{web-hosted} (managed by a third-party service provider), or \emph{locally-deployable} (managed by the enterprise using the platform). IoT management platforms are different from other web management platforms in many aspects~\cite{zhou2021reviewing}. For example, these platforms obtain data from multiple sources (devices, SIM cards, users), and each of them could become a potential source of malicious data entry. Also, these platforms manage real world IoT deployments. As many of them support critical operations in industrial and enterprise IoT setups, exploiting vulnerabilities in them could have severe physical world consequences in addition to revenue loss and day to day business disruption.

Existing work demonstrated several attacks on cellular IoT that could cause overcharging through voice services~\cite{xie2018voice}, and create financial losses for the cellular operators by using IoT SIMs inside non-IoT devices~\cite{xie2020can}. 
Connectivity management services are a major component of cellular IoT for SIM management tasks. Apart from a recent BlackHat presentation~\cite{blackhat} (conducted on 9 platforms), there is no comprehensive evaluation of vulnerabilities in these services.
Unauthorized access control issues in IoT platforms have been studied in the past~\cite{li2020iot} by analyzing IoT devices' mobile companion applications. However, such studies have only covered the API endpoints called from the client-side, i.e., management APIs with more serious security issues \textcolor{black}{remaining} unexplored. Apart from web-hosted platforms, there are also widely-used locally-deployable solutions (both open and closed-source) for IoT deployment, which so far have not been analyzed.

We design and implement a generalized security framework to evaluate the security posture of IoT management platforms from an external attacker's perspective, 
focused on the key services provided by them---i.e., connectivity, device, and data management. 
Our evaluation framework comprises a wide range of vulnerabilities such as broken authentication, unauthorized access, vulnerable trigger-action function and lack of input validation.
For the evaluation, %
we rely on a combination of automated, semi-automated and manual vulnerability detection techniques; for web-hosted platforms these tests are carefully applied not to interfere with the platform operations. The scope of evaluation comprises the web requests generated upon using the platforms' websites, and those corresponding to the platforms' standalone APIs. We use custom Python scripts to simulate the behavior of IoT devices making API calls to the platforms. We also use virtual machines (VMs) to mimic Linux-based IoT devices for evaluating platforms that provide agent software for device management. To evaluate locally-deployable platforms, we install their Linux-compatible versions inside VMs. %

We used our framework to conduct a security evaluation of 52 IoT management platforms (42 web-hosted and 10 locally-deployable), and found major security weaknesses in several platforms. %
Vulnerabilities detected in them could impact multiple stakeholders---the platform itself, the enterprise users, the end consumers, and all the devices connected to these platforms, with consequences such as: %
disconnecting IoT devices from
the cellular network, sending arbitrary unauthorized commands to devices, and disclosure of device metadata including
GPS coordinates. Moreover, poorly configured devices can be leveraged for platform-level attacks against the corresponding platforms and their users; e.g., authentication tokens/keys intercepted
from the IoT devices that use HTTP/MQTT (without TLS) can be leveraged
by an attacker to perform cross-site scripting (XSS) via forged data
submissions. 

\subhead{Contributions and notable findings} 
\begin{enumerate}
\item We design a comprehensive security evaluation framework for evaluating various complex functionalities offered in modern IoT management platforms. We include tests pertinent to core platform services---connectivity, device, and data management for operating a large number of IoT devices. 
We consider several practical attacker models, including a regular remote attacker, an on-path attacker, and an attacker requiring minor user involvement (e.g., clicking on an attacker provided link). We realize this framework using a carefully-deployed combination of existing tools (in addition to our own scripts), as our tests are performed on live services (albeit on our own test accounts) along with locally-deployable platforms.

\item We apply our framework on 52 selected IoT management platforms of various sizes, offering a wide-range of services. 
Our analysis uncovered vulnerabilities in 33/52 platforms. 17/52 platforms were affected by vulnerabilities that could lead to platform-wide attacks affecting all users and all connected devices.
This indicates that our framework is both \emph{applicable} (i.e., can handle varying, complex services), and \emph{effective} (i.e., can detect serious security problems).

\item The unauthorized access vulnerabilities that we found in 9/52 IoT platforms, involve key platform features that can be abused to launch serious attacks against service availability, reliability, and billing. Such attacks include: arbitrary SIM deactivation, unauthorized Short Message Service (SMS) delivery and forged data submission from IoT devices. 

\item 13/52 platforms are affected by broken authentication, with varying implications on their services---e.g., full account takeover of any user (in Aeris Neo, AskSensors, Favoriot, MDash, ResIOT, Fogwing, Thingsboard, GlobalM2MSIM), and preventing IoT SIM registration and activation (Hologram, KeepGo and OneSIMCard).

\item A vulnerable trigger-action function (cf.~\cite{ahmadpanah2021}) in
\mbox{\url{TheThings.io}} grants root access to a Kubernetes container shared across users on the platform, by breaking out of their JavaScript sandbox. Missing sandbox implementation for trigger-action function in OpenRemote grants root access to the server to any platform user.  %

\item
16 platforms lack proper input validation checks, making them all 
vulnerable to XSS attacks, and one (ResIOT's locally-deployable platform) vulnerable to SQL injection. On 9 platforms, an adversary can abuse XSS to steal session credentials (browser cookies and login tokens). %
\item \textcolor{black}{We were assigned 11 common vulnerabilities and exposures (CVEs) for vulnerabilities uncovered in locally-deployable platforms (see Table~\ref{table:cves} in Appendix~\ref{supp-info}).} Among these, one CVE was rated as critical (score 9.8/10), 3 CVEs rated as high severity (2 CVEs with score 8.8/10 and one CVE with score 7.2/10), and 7 CVEs rated as medium severity (with scores between 4.3/10 and 6.5/10). 
We also received monetary awards as bounty from two companies (Boodskap and Platform X\footnote{Actual name withheld according to the guidelines of the company's bug disclosure program. Throughout the article, this company will be referred to as Platform X.}) along with a certificate of appreciation from Verizon.

\end{enumerate}

\subhead{Ethical considerations and responsible disclosure} 
We performed all the tests on our own accounts. For inadvertent access to sensitive data (e.g., authentication credentials in error messages), as per our university's ethics guidelines, we informed the affected platform in a timely fashion and did not retain the data.
We did not perform any active scanning via automated tools for vulnerability detection and exploitation, 
to avoid any adverse effect on the day-to-day usage of the web-hosted platforms. For sandbox breakout on TheThings.io, we followed a coordinated disclosure process, and we were granted a complimentary paid account by the platform for comprehensive testing. %
We reported our findings to all the affected platforms via emails/support tickets. In one instance (Asksensors), upon not receiving any response, we also took the help of the CERT-FR (\url{cert.ssi.gouv.fr}). 
At the time of this submission, we received responses from 28 platforms.  
Aeris Neo took down their portal for fixing. RemoteIOT promptly acted on our disclosure and remediated the reported vulnerability within a day (which we also confirmed). KeepGo, Favoriot, Fogwing, SocketXP and ResIOT have also resolved the reported issues. Asksensors and Verizon claimed to have fixed all issues, but on a second inspection, we found some fixes to be inadequate, and as such informed the companies again. 6 other platforms applied partial fixes. 
Another 9 platforms indicated that they are working on the fixes. 4 platforms (Telnyx, Pelion, Tago, SIMControl) where minor security issues were reported, responded that they consider those issues as acceptable risks.

\noindent\textcolor{black}{\subhead{Differences with the CODASPY version~\cite{btcodaspy}}
This article is an extended version of a 6-page conference paper~\cite{btcodaspy}. The main addition in this article is a comprehensive security evaluation of 10 locally-deployable platforms, including 8 open-source and 2 closed-source solutions. We provide details of the vulnerabilities found in locally-deployable platforms and the CVE-IDs assigned for them. We also include five additional vulnerabilities in our testing framework: cross-site request forgery, insecure communication, misconfigured cookie attributes, information disclosure via error messages, and information leakage to third parties. The article also contains an additional set of results for the four main vulnerabilities discussed in~\cite{btcodaspy}. We provide a detailed update on the responses received from the affected platforms to our responsible disclosure. Finally, this article also includes key insights from our analysis and recommendations for platform developers and users.}

\section{Background and Threat Model}
\label{section:background}
\noindent In this section, we summarize key functionalities offered by IoT management services, and provide our threat model.

\subhead{IoT connectivity management}
With the widespread deployment of 4G/5G technologies, cellular connectivity is becoming the preferred option for many consumer/industrial IoT devices. %
Although the IoT SIM cards (also called as programmable wireless SIM cards, Machine to Machine/M2M SIM cards) use the same network as the cellular network, there are some notable differences. Firstly, IoT SIM cards are typically offered for global connectivity via multiple carriers. Secondly, the IoT SIM owners %
can manage their SIM cards via connectivity management services, e.g., to register, activate, pause, track usage, change subscription plans, and decommission devices. These tasks can be performed automatically at scale, through web APIs provided by the connectivity services. 

IoT SIM cards can typically be purchased from the providers' websites after creating a user/business account on the portal; some providers do not sell SIM cards to individual users, and involve a manual verification process. %
Following is a brief outline of each state in the IoT SIM card's lifecycle~\cite{emnify-sim}: when the SIM card is delivered to the user, it remains at the \emph{initial state} (not connected to the cellular network);
after registering a card on the portal (e.g., by entering an activation code printed on the card), it reaches the \emph{active state} and can connect to the cellular network, and the user is billed for the SIM usage;
a temporarily suspended card is in the \emph{paused state} (unable to access the cellular network); and 
a permanently disconnected card from the cellular network is in the \emph{terminated state} (e.g., when an IoT device is decommissioned).

Key features offered by the connectivity management services for enterprise users include:
managing the SIM state %
and connection troubleshooting in real-time;
setting usage limits on data consumption for each SIM card, or a group of SIM cards;
sending/receiving SMS messages to/from the SIM card (e.g., for commands/outputs);
setting the International Mobile Equipment Identity (IMEI) lock to prevent abuse of stolen cards;
generating reports on data usage and billing; and
creating rules to generate alerts in case an anomalous behavior is detected, e.g., exceeding data consumption limit.

\subhead{IoT device management}
Enterprises typically deal with a large number of devices and these deployments need to be managed remotely. IoT device management services offer a centralized web portal to perform such key administrative tasks on the IoT devices, as well as web APIs to enable automation. Key features in device management include~\cite{devicemgmtdef}:
provisioning and authentication of devices (devices are assigned unique IDs, and authentication tokens, which are then used for onboarding on the platform);
adjusting the configuration of the IoT device as needed (e.g., adding a new variable in the JSON object sent to the platform, modifying the API's URL),
monitoring usage, and diagnostics for troubleshooting; and  allowing enterprises to upload software/firmware updates, which are subsequently pushed to IoT devices.

\subhead{IoT data management}
IoT data management services enable the centralized data aggregation and processing for IoT devices. Similar to device management, the IoT device is onboarded on the platform with a unique device ID and authentication token. Data is typically submitted to the platform via protocols such as HTTP, MQTT, AMQP and COAP~\cite{naik2017choice}. %
Users can also create visualization dashboards
based on IoT device data for analytics purposes.

\subhead{Threat model}
The scopes of the considered attacks are defined as follows. A \textit{platform-wide attack} affects all users and all connected devices of the platform; examples include: remote code execution via sandbox escape (platform infrastructure compromise), attacks involving broken authentication and unauthorized access issues that require the use of easily-enumerated identifier values (e.g., short numeric IDs, sequential SIM numbers). A \textit{user-specific attack} can affect only a specific user and the devices owned by that user; examples include: session hijacking via \textcolor{black}{cross-site scripting (XSS)}, password reset via \textcolor{black}{cross-site request forgery (CSRF)}, user credential theft via \textcolor{black}{secure sockets layer (SSL) stripping}, attacks involving broken authentication and unauthorized access vulnerabilities that require specific user/device IDs that are not easily-enumerated/guessed (e.g., UUIDs, registered email address of the target user). A \textit{device-specific attack} can affect a specific device (e.g., intercepting HTTP traffic to steal device authentication credentials). 
To perform these attacks, we assume three types of attackers in our threat model. A \emph{user-independent remote attacker} directly interacts with the platform and does not need to involve the victim user/device in any manner. Such an attacker can create a user account on the platform's website, and perform the intended attacks. %
A \emph{user-dependent remote attacker} also performs the attack remotely, but requires user involvement (such as clicking on a phishing URL).
An \emph{on-path attacker} must be on the same network path as the victim user/IoT device, and collect and analyze the traffic flow between the  user's browser/IoT device and the platform's server.
Attacks requiring physical access to an IoT/user device are out of scope. %

\section{Security Analysis Framework}
\label{section:framework}
\begin{figure}[t]
    \includegraphics[height=.7\textheight]{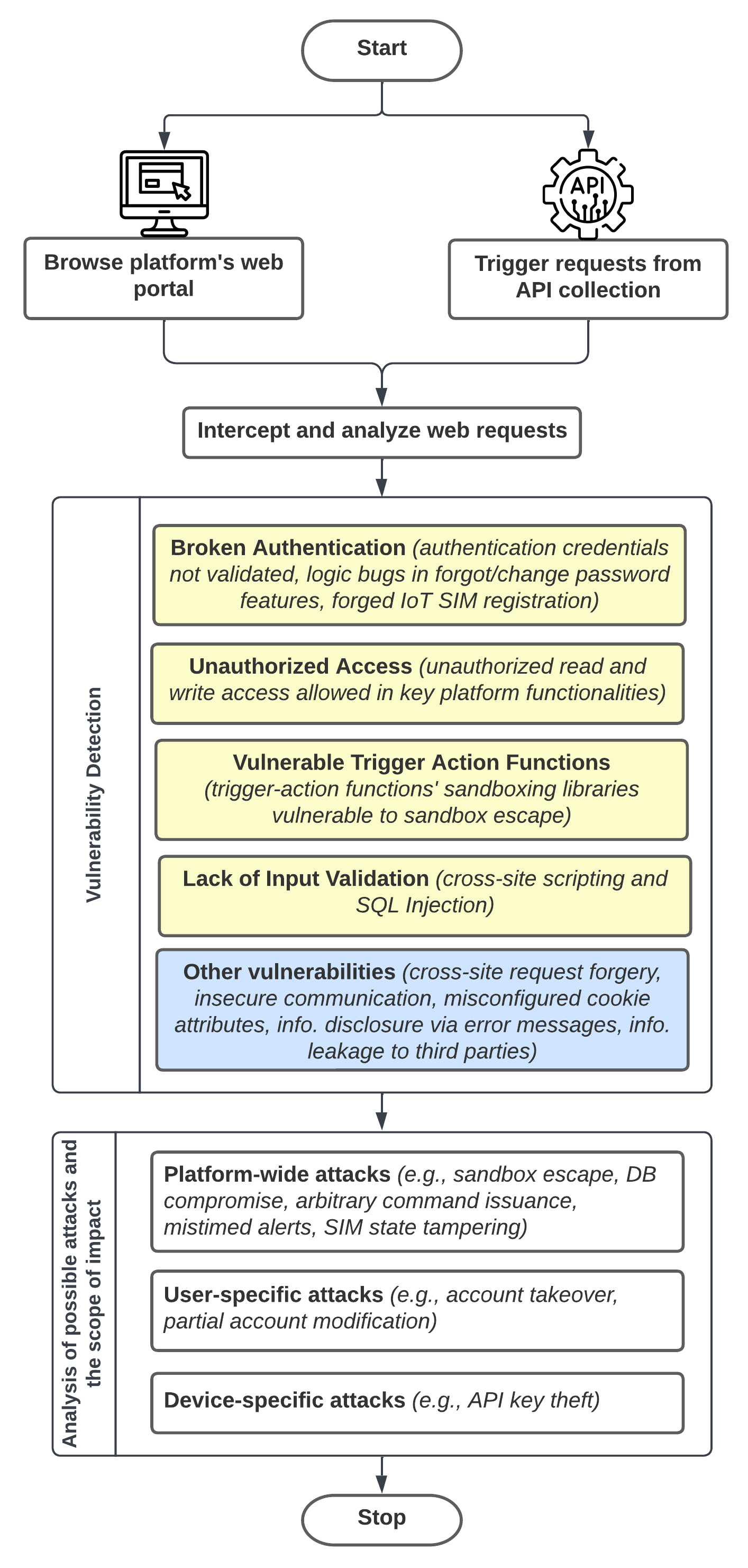}
    \caption{Overview of proposed security analysis framework. Yellow box indicates that vulnerability detection requires active interaction with the IoT management platform's web server; the blue box ones can be detected by passively monitoring the web traffic between the browser and the platform. The attacks can result from one or multiple vulnerabilities.}
    \label{fig:framework}
\end{figure}

\noindent In this section, we provide a framework for performing the security analysis of IoT management platforms. %
At first, we identify key platform functionalities by manually analyzing a few platforms and perusing their API documentations. 
Then we select an initial list of potential security vulnerabilities
that is motivated by prior research in the field of IoT security~\cite{alrawi2019sok, li2020iot, fernandes2016security,obermaier2016analyzing, ahmadpanah2021, wang2018cracking, wang2019looking, 
jiang2020experimental, 
zhou2021reviewing, wang2019charting, chen2019your, rondon2021lightningstrike}, and online services security~\cite{barth2008robust, 
drakonakis2020cookie, gadient2020web}. 
We then iteratively refine the list of associated vulnerabilities based on their impact on the key functionalities,
and focus on the vulnerabilities applicable to multiple platforms. Figure ~\ref{fig:framework} provides an overview of the proposed framework. In this section, we discuss the process of detection of each vulnerability along with the potential impact on the affected platforms, users, and connected devices. For each platform, we create two user accounts, and use one as an attacker and the other as a victim user.
We also check for some other security and privacy issues (detailed in Appendix~\ref{section:othervulns}) that can amplify/augment the primary attacks discussed in this section. 

\subsection{Broken Authentication}
 Broken authentication~\cite{wang2019looking, jiang2020experimental, zhou2021reviewing} in IoT management platforms can lead to platform-wide attacks such as SIM state tampering, device data tampering, user information disclosure, arbitrary command delivery, and firmware theft. Broken authentication could also lead to user-specific attacks such as account takeover, which grants an attacker read and write access to all functionalities accessible to the targeted user. We perform the following checks to detect insecure implementation of authentication. %

    We log in to the platform and capture all the web requests that require authentication (e.g., cookies and API keys). We also capture requests by issuing API requests from standalone API collections (if provided).  
    We exclude web requests for loading static content such as JavaScript files and images. We then remove the session credentials from each request and resend it to the platform; we label the platform as vulnerable if the response for any modified web request is the same as the original one.
    We use Auth Analyzer~\cite{portswiggerauth} for these checks, followed by manual review to assess the impact.

    We check for logic bugs in the forgot password and password reset functionalities of the platforms' websites. We test if it is possible to reset the password of another user by tampering parameters (e.g., email address, username) in the underlying web requests.

    After purchasing an IoT SIM card (where possible), to activate the card and manage its lifecycle, the card owner must create an account on the platform and add the card to their platform account. We check if the platform validates whether the user owns the card that they are trying to add to their account. Absence of such validation 
    could allow forged IoT SIM registration, wherein an attacker is able to add an unassigned card to their account, which when purchased by a legitimate user, would fail to register, and may require intervention from the platform's technical support.

\subsection{Unauthorized Access}
Unauthorized access~\cite{li2020iot, zuo2017authscope, chen2019your, drakonakis2020cookie} refers to the inability of an IoT management platform to validate if the entity requesting a resource is allowed to access that resource, and to what extent. These vulnerabilities could result in platform-wide attacks such as SIM state tampering, device data tampering, arbitrary SMS messages, user information disclosure, alerts configuration leakage, SMS message leakage and device data leakage. 
We perform the following actions to identify such issues. 
(a) We create two user accounts for each target platform -- one as the victim account, and the other as an attacker account. %
(b) We log in with the victim account on the platform, and
capture the underlying web request to test for unauthorized access. 
(c) We replace the authentication details in the web request captured in step (b) with those of the attacker account, and send the modified request to the platform.
(d) We observe the response of the modified request sent from the attacker's account. If the modified request is successfully processed, we flag the platform as vulnerable to unauthorized access. We use Auth Analyzer to detect unauthorized access vulnerabilities, followed by manual review of the affected requests to 
assess the impact. For this vulnerability, the platforms do check for the authentication credentials in the web requests, but fail to validate if the requesting users have the permission to access the resources.
Unauthorized access can lead to several abuse scenarios in the following key functionalities of device, data and connectivity management services.

\subhead{IoT SIM management}
A user needs to manage the IoT SIM card through different state changes (activation, deactivation, pause) in its lifecycle, which must be properly access-controlled. 
Unauthorized activation of SIM cards in non-functional IoT devices can lead to unnecessary subscription charges, and unauthorized deactivation of a  card would disconnect the corresponding IoT device. %

\subhead{SIM and service alerts} A user can set up notification alerts 
to limit %
data usage for IoT SIM cards. A user can also write custom trigger functions for conditional alerts based on data transmitted by the devices. By abusing unauthorized read access to these alert settings, an attacker could infer the nature of data transmitted by the devices and could obtain sensitive information such as the API keys for third-party services used in the trigger-functions.
Unauthorized write access allows an attacker to change the notification rules for the user's SIMs/devices, affecting day-to-day operations and causing financial losses. %

\subhead{SMS}
SMS messages are widely used for managing configuration of IoT devices and for sending and receiving 
data from IoT devices~\cite{pistek2020using}. 
Unauthorized read access could allow an attacker to obtain sensitive information exchanged between IoT devices and the platform via SMS, whereas unauthorized write access can be abused to send SMS messages to users' IoT SIMs, which may lead to arbitrary command execution in IoT devices and overcharging of the victim's account.
    
\subhead{IMEI lock}
Users can set up an IMEI 
lock for their IoT SIM cards by specifying the IMEI number of the devices whitelisted for using the SIM card. 
This helps to prevent the misuse of stolen IoT SIM cards. 
Unauthorized IMEI lock could allow an attacker to %
interrupt the cellular connectivity for the affected card, %
disable the IMEI lock protection altogether, and use the IoT SIM card outside the designated IoT devices, leading to financial loss for the card owner.  

\subhead{Remote commands} Users can remotely execute commands on their IoT devices via the platform's website. Insufficient access control %
may allow an attacker to execute arbitrary commands on target devices. Some platforms also store the output of the executed commands, for later viewing by the user. If unprotected, potentially sensitive information may be exposed from these outputs. %

\subhead{Firmware updates} 
Users can upload firmware files for their IoT devices on the platform's website. These files can also be downloaded from the platform. By abusing unauthorized read access, an attacker can download files uploaded by other users, which may cause breach of intellectual properties~\cite{zhou2019discovering}. On the other hand, unauthorized write access could be abused to inject malicious firmware in targeted IoT devices.

\subhead{IoT device data}
Platforms must ensure that an IoT device's data is available only to the device owner. %
A platform also must validate incoming data submission requests to confirm if the requests are coming from a legitimate IoT device, e.g., via checking access tokens. Unauthorized read access to these tokens could be abused by an attacker to submit forged data from the affected IoT devices. %

\subhead{Account information} Users provide \textcolor{black}{personal identifiable information (PII),} e.g., user name, email ID, mobile phone number, address, and organization name, during registration of an account. %
Unauthorized read access can leak such PII,
and unauthorized write access in functionalities such as change password could allow an attacker to take over a user's account.

\subsection{Vulnerable Trigger-Action Functions}
\label{vuln_trig}
Trigger-Action Platforms (TAPs~\cite{ahmadpanah2021,wang2019charting}) connect IoT devices and cloud services with the help of trigger-action applications. When a trigger is received by the application, certain actions are performed, which are programmed into the trigger-action application. Some IoT management platforms offer similar trigger-action features on their websites, allowing users to write code for custom functionality (typically in JavaScript), e.g., unit conversion, and periodic tasks. %
Custom functions must be written for parsing and processing data payloads received from IoT devices; see \cite{triggerexample} for an example of such a function which triggers an email when the temperature reading received from a device goes beyond a set limit.

Platforms use sandbox libraries with limited set of JavaScript methods to securely execute these user-supplied trigger-functions in an isolated environment, and to avoid attacks such as remote code execution. 
Popular NodeJS libraries 
are known to have sandbox bypass vulnerabilities and can be exploited in IoT communication (cf.~\cite{ahmadpanah2021} for TAPs vulnerabilities on IFTTT and Zapier platforms). 
We first try to find out the sandbox library used by the platform through a stack trace using the following JavaScript code: 
{\small{\verb!function main(params, callback)!}}
{\small{\verb!{callback(new Error().stack);}!}}

If the detected JavaScript sandboxing library has known vulnerabilities, we check for those vulnerabilities, and use benign system commands such as \texttt{id} for confirming the privilege level of the system access granted by the vulnerability. We follow a process of coordinated disclosure with affected platforms to minimize any accidental system impact.

\subsection{Lack of Input Validation}
IoT management platforms must validate the inputs present in the %
web requests before processing them. Also, any data received from IoT devices must be suitably encoded before displaying on the platform's website. The lack of input validation~\cite{jiang2020experimental, alrawi2019betrayal}
enables attacks, e.g., XSS and SQL injection. While XSS can lead to user-specific attacks, e.g., account takeover, SQL injection can cause platform-wide database compromise. We detect the presence of these vulnerabilities by sending web requests with malformed input parameters and analyzing the corresponding web responses. To detect XSS, we provide custom JavaScript payloads (e.g.,
{\small{\verb!<script>alert(1)</script>!}})
in the input parameters, and check if the supplied payload is executed in the browser while navigating the website.
Similarly, we checked if SQL errors are returned upon appending a single quote at the end of input parameter values.
Due to ethical/legal concerns, we do not use automated scanning tools to flood the platforms with requests containing malicious inputs. All instances of this vulnerability on web-hosted platforms have been solely detected via manual inspection. Burp's active scanner is used to supplement manual testing on locally-deployable platforms. 

\subsection{Other Vulnerabilities}
\label{sec:othervulnsdesc}
\textcolor{black}{We also check for some other security issues as follows (detailed in Appendix~\ref{section:othervulns}). 
(1) Cross-site request forgery (CSRF): We check if a user-dependent remote attacker can tamper with the victim user's assets (user account and underlying IoT SIMs/devices) by exploiting missing CSRF protection in key platform functionalities.
(2) Poorly configured cookie attributes: We check if the Secure and HTTPOnly attributes are set for the session related cookies. Absence of these attributes, coupled with XSS/insecure communication could be exploited by a user-dependent/on-path attacker respectively to takeover a target user's account.
(3) Insecure communication: We check if the platform allows an on-path attacker to passively capture device API keys (upon using HTTP/MQTT without TLS), or collect login password (via SSLStrip).
(4) Information disclosure via error messages: While testing for lack of input validation, we also check if the platform returns verbose error messages, and check if any sensitive information is disclosed in the error messages, as they might help an attacker know more about the target platform's architecture. 
(5) Information leakage to third parties: While browsing a platform's website, we check if any sensitive information is sent to a third-party.}

\textcolor{black}{The first three issues can be abused along with lack of input validation to perform chained attacks with amplified impact. The fourth issue can be abused by any remote attacker for platform reconnaissance. The last issue pertains to privacy leakage on IoT platforms.}

\section{Target Platforms and Analysis Setup} %
\label{section:setup}
\textbf{Target platforms}. Since the IoT platforms are used mostly by enterprise users/developers and not meant for mass consumption, typical website ranking services (such as Tranco\footnote{\url{https://tranco-list.eu}}) cannot be used for platform selection. Instead, we rely on
a combination of the following sources to gather the list of 52 web-hosted/locally-deployable platforms: (a) 
survey papers on IoT platforms~\cite{ray2016survey, hejazi2018survey,yu2019analysis, zdravkovic2016survey}; (b) top search engine results with keywords such as IoT platform, IoT device management, IoT data management, IoT connectivity, IoT SIM, M2M SIM. The selected web-hosted platforms cater to a large number of IoT devices (self-reported by the platforms, as of Sept.\ 1, 2022): e.g., Verizon (31M), Telenor (17M), Emnify (10M), RemoteIOT (10M). For the selected locally-deployable platforms, 
we found (as of Sept.\ 1, 2022) several instances from Censys.io, e.g., Thingsboard (2206), DGIOT (130), OpenRemote (57); note that these counts include only the internet exposed deployments, and each deployment may serve many IoT devices.

30 platforms allowed creation of trial enterprise user accounts, in which we registered using fictitious company names such as \textit{TestCompany}. For the rest, we registered for trial accounts as individual users. 
20 platforms (all web-hosted) we assessed offer connectivity management service via cellular networks. 
We purchased IoT SIM cards from 6/20 platforms, namely, Telnyx, Telenor, Hologram, OneSIMCard, KeepGo and OpenM2M. For these 6 platforms, we performed the security analysis with the privileges of valid enterprise users of these platforms. For the remaining 14/20 platforms offering connectivity management services, and for all the platforms analyzed that offer device and data management services, we conducted the analysis with the trial user accounts, typically having access to a limited set of functionalities.
In addition to 42 web-hosted platforms, we also evaluated 10 locally-deployable IoT platforms, which includes 2 closed-source and 8 open-source platforms. 

\subhead{Analysis setup} %
We perform a black-box security assessment of the platforms and analyze their web applications and web APIs. We use Burp Suite~\cite{portswiggerburp} as a man-in-the-middle proxy to intercept the web requests and send crafted web requests to the platforms. 
\begin{table}[]
\centering
\caption{List of analysis tools and approaches}
\scalebox{0.8}{
\begin{tabular}{lll}
\hline
\textbf{Security Vulnerability}                                                & \textbf{Analysis Type} & \textbf{Tool/Technique}   \\ \hline
Broken authentication                                                            & Semi-automated         & Auth Analyzer~\cite{portswiggerauth}      \\ 
Unauthorized access                                                          & Semi-automated         & Auth Analyzer~\cite{portswiggerauth}      \\ 
\begin{tabular}[c]{@{}l@{}}Vulnerable trigger-action functions\end{tabular} & Manual                 & -                  \\ 
Lack of input validation                                                       & Manual                 & -                  \\ 
Cross-site request forgery                                                     & Semi-automated         & CSRF Scanner~\cite{csrfscan}       \\ 
Insecure communication                                                         & Automated         & Bettercap~\cite{bettercap}          \\ 
  Information disclosure via error messages &
  Automated &
Error Message Checks~\cite{errorscan} \\ 
Misconfigured cookie attributes                                                    & Automated         & Burp Suite         \\ 

  \begin{tabular}[c]{@{}l@{}}Info. leakage to third parties\end{tabular} &
  Semi-automated &
  \begin{tabular}[c]{@{}l@{}}Regex based search\end{tabular} \\ \hline
\end{tabular}
}
\label{table:tools-used-main}
\end{table}

We use the following approaches in our analysis (see Table~\ref{table:tools-used-main}). 
We perform manual testing to check lack of input validation and vulnerable trigger-action functions; no automated scanning tools are used to avoid affecting the platform operations. For trigger-action functions, we manually explore the available sandbox-escape attacks for a given library.  
The following tests are semi-automated: broken authentication and unauthorized access (manually exploring key functionalities on the website to capture underlying web requests), CSRF (removing false positives reported by CSRF Scanner~\cite{csrfscan}),  information leakage to third parties (tuning the regular expressions for each platform). 
Finally, information disclosure via error message, misconfigured cookie attributes, and testing for insecure communication are performed automatically. Impact analysis is done manually. %

To mimic the behavior of real IoT devices, we used Curl\footnote{https://github.com/curl/curl} and custom Python scripts to issue test HTTP requests to the platforms. We needed to install an edge agent on an IoT device to perform the analysis on 10/52 IoT platforms. We installed Linux-compatible versions of these agents inside virtual machines. To evaluate the locally-deployable IoT platforms, we set up each platform inside a separate VM. %
While testing these local deployments, we used Burp Suite's active scanner in conjunction with manual testing. Since we perform a black-box assessment, we do not conduct a source code review for the open-source locally-deployable platforms.

\begin{table*}[]
\centering
\caption{\textcolor{black}{Overview of the discovered attacks, their scope, corresponding vulnerable platform functionalities and
underlying vulnerabilities: Broken authentication (\faUnlock), Unauthorized access (\faUserTimes), Vulnerable trigger action functions (\faCode), Lack of input validation (\faKeyboard), Cross-site request forgery (\faStar), Insecure communication (\faExclamationTriangle), Misconfigured cookie attributes (\faStop). Notations seperated by ~$\parallel$~ indicate that the attack is possible by exploiting either of the denoted vulnerabilities. Notations seperated by + indicate that the attack requires a combination of vulnerabilities to be exploited.}
}
\resizebox{\textwidth}{!}{
\begin{tabular}{llcccccccc}
\hline
\multirow{2}{*}{\textbf{\begin{tabular}[c]{@{}l@{}}Attack \\ Scope\end{tabular}}} & \multirow{2}{*}{\textbf{Attack Description}} & \multicolumn{8}{c}{\textbf{Vulnerable Platform Functionalities}}                                                                                                                                                                                                                                                                                                                                                                                                                                                                                                                                                                                                  \\ \cline{3-10} 
                                                                                  &                                              & \multicolumn{1}{l}{\textbf{\begin{tabular}[c]{@{}l@{}}SIM\\ Mgmt.\end{tabular}}} & \multicolumn{1}{l}{\textbf{\begin{tabular}[c]{@{}l@{}}Alerts\\ Mgmt.\end{tabular}}} & \multicolumn{1}{l}{\textbf{SMS}} & \multicolumn{1}{l}{\textbf{\begin{tabular}[c]{@{}l@{}}IMEI \\ Lock\end{tabular}}} & \multicolumn{1}{l}{\textbf{\begin{tabular}[c]{@{}l@{}}Device \\ Commands\end{tabular}}} & \multicolumn{1}{l}{\textbf{\begin{tabular}[c]{@{}l@{}}Firmware \\ Updates\end{tabular}}} & \multicolumn{1}{l}{\textbf{\begin{tabular}[c]{@{}l@{}}Device Data\\ Handling\end{tabular}}} & \multicolumn{1}{l}{\textbf{\begin{tabular}[c]{@{}l@{}}User \\ Accounts\end{tabular}}} \\ \hline
\multirow{11}{*}{Platform-wide}                                                   & Sandbox escape                               &                                                                                  & \faCode                                                                                   &                                  &                                                                                   &                                                                                         &                                                                                          &                                                                                      &                                                                                       \\
                                               & Database compromise                               &                                                                                  &                                                                                  &                                  &                                                                                   &                                                                                         &                                                                                         &                                                                                    &                      \faKeyboard ~+~\faStar~~~~~~~~~~~~~~~~~~~~~                                                            \\
                                                                  & Arbitrary command issuance                   &                                                                                  &                                                                                     &                                  &                                                                                   & \faUnlock                                                                                       &                                                                                          &                                                                                      &                                                                                       \\
                                                                                  & Mistimed alerts                       &                                                                                  & \faUnlock~$\parallel$~\faUserTimes                                                                                   &                                  &                                                                                   &                                                                                         &                                                                                          &                                                                                      &                                                                                       \\
                                                                                  & SIM state tampering                          & \faUnlock~$\parallel$~\faUserTimes                                                                                &                                                                                     &                                  & \faUserTimes                                                                                 &                                                                                         &                                                                                          &                                                                                      &                                                                                       \\
                                                                                  & Device data tampering                        &                                                                                  &                                                                                     &                                  &                                                                                   &                                                                                         &                                                                                          & \faUnlock~$\parallel$~\faUserTimes                                                                                    &                                                                                       \\
                                                                                  & Arbitrary SMS messages                       &                                                                                  &                                                                                     & \faUserTimes                                &                                                                                   &                                                                                         &                                                                                          &                                                                                      &                                                                                       \\
                                                                                  & User information disclosure                  &                                                                                  &                                                                                     &                                  &                                                                                   &                                                                                         &                                                                                          &                                                                                      & \faUnlock~$\parallel$~\faUserTimes~~~~~~~~~~~~~~~~~~~~~                                                                                      \\
                                                                                  & Alerts config. leakage                       &                                                                                  & ~\faUserTimes                                                                                    &                                  &                                                                                   &                                                                                         &                                                                                          &                                                                                      &                                                                                       \\
                                                                                  & SMS message leakage                          &                                                                                  &                                                                                     & ~\faUserTimes                               &                                                                                   &                                                                                         &                                                                                          &                                                                                      &                                                                                       \\
                                                                                  & Firmware theft                               &                                                                                  &                                                                                     &                                  &                                                                                   &                                                                                         & ~\faUserTimes                                                                                        &                                                                                      &                                                                                       \\
                                                                                  & Device data leakage                          &                                                                                  &                                                                                     &                                  &                                                                                   &                              ~\faUserTimes                                                          &                                                                                          & \faUnlock~$\parallel$~\faUserTimes                                                                                    &                                                                                       \\ \hline
\multirow{2}{*}{User-specific}                                                    & Account takeover                             &                                                                                  &                                                                                     &                                  &                                                                                   &                                                                                         &                                                                                          &                                                                                      & \faUnlock~$\parallel$~\faStar~$\parallel$~(\faKeyboard~+~\faStop~+~\faExclamationTriangle)                                                                                     \\
                                                                                  & Partial account modification                 &                                                                                  &                                                                                     &                                  &                                                                                   &                                                                                         &                                                                                          &                                                                                      & \faStar~~~~~~~~~~~~~~~~~~~~~~~~~~~~                                                                                     \\ \hline
Device-specific                                                                   & API key theft                                &                                                                                  &                                                                                     &                                  &                                                                                   & \faExclamationTriangle                                                                                      & \faExclamationTriangle                                                                                       & \faExclamationTriangle                                                                                   &                                                                                       \\ \hline
\end{tabular}
}
\label{tab:attacksinkeyfunctionalities}
\end{table*}

\section{Results}
\label{section:results}
We use our framework to analyze the security posture of 52 IoT management platforms. The analysis was done between January 2021 and June 2022. \textcolor{black}{We summarize the attack types and underlying vulnerabilities that are affecting various functionalities in Table~\ref{tab:attacksinkeyfunctionalities}.}
In this section, we provide a detailed discussion on the findings of our analysis; Table~\ref{tab:shortlistedplatformvulns} provides an overview of the  findings based on the discovered instances of data exposure and malicious write 
along with the type of attacker and the scope of attack.

\begin{table*}[!htb]
\centering
\caption{Overview of the discovered instances of data exposure and malicious write along with the type of attacker and the scope of attack (see Sec.~\ref{section:background}). 
Any remote attacker can perform platform-wide (\redcirc) or user-specific (\fullcirc) attacks; a user-dependent remote attacker can perform platform-wide (\purplecirc) or user-specific (\emptycirc) attacks; any on-path attacker can perform user-specific (\faTimesCircle[regular]) or device-specific (\faDotCircle[regular]) attacks. 
In instances exploitable by multiple attacker types, we consider the worst one (e.g., remote attackers are worse than on-path attackers), with the broadest scope.
Platform notation: web-hosted if not mentioned, (O): open-source locally-deployable, (C): closed-source locally-deployable.} %

\begin{tabular}{llllll|lllllllllll}
                       & \multicolumn{5}{c|}{\textbf{Data Exposure}}                                                                                    & \multicolumn{11}{c}{\textbf{Malicious Write}}                                                                                                                                                                                                                                     \\ \cline{2-17} 
\textbf{}              & \rotatebox[origin=l]{90}{\textbf{Alerts Config.}}
 & \rotatebox[origin=l]{90}{\textbf{SMS Messages}}
 & \rotatebox[origin=l]{90}{\textbf{Device Metadata}}
 & \rotatebox[origin=l]{90}{\textbf{Command output}}
 & \rotatebox[origin=l]{90}{\textbf{Account Info}}
 & \rotatebox[origin=l]{90}{\textbf{SIM State}}
 & \rotatebox[origin=l]{90}{\textbf{SIM Registration}}
 & \rotatebox[origin=l]{90}{\textbf{Alerts Config.}}
 & \rotatebox[origin=l]{90}{\textbf{SMS Messages}}
 & \rotatebox[origin=l]{90}{\textbf{IMEI Lock}}
 & \rotatebox[origin=l]{90}{\textbf{Device Commands}}
 & \rotatebox[origin=l]{90}{\textbf{Forged IoT data}}
 & \rotatebox[origin=l]{90}{\textbf{Dashboard}}
 & \rotatebox[origin=l]{90}{\textbf{Account Takeover}}
 & \rotatebox[origin=l]{90}{\textbf{DB Modification}}
 & \rotatebox[origin=l]{90}{\textbf{Potential RCE}}
 \\ \hline
Verizon’s   Thingspace &                         &                       &                            &                         &                       &                    &                             &                        &                       &                    &                          &                          & \emptycirc                &                           &                          &                        \\
Platform X          &                         &                       & \redcirc                        & \redcirc                     &                       &                    &                             &                        &                       &                    &                          &                          & \fullcirc                &                           &                          &                        \\
Aeris Neo              & \redcirc                     & \redcirc                   &                            &                         & \redcirc                   & \redcirc                &                             & \redcirc                    & \redcirc                   &                    &                          &                          &                    & \redcirc                       &                          &                        \\
RemoteIOT              &                         &                       &                            &                         &                       &                    &                             &                        &                       &                    &                          &                          & \fullcirc                &                           &                          &                        \\
OneSIMCard             &                         & \redcirc                   &                            &                         &                       & \redcirc                & \redcirc                         &                        & \redcirc                   & \redcirc                &                          &                          &                    &      \emptycirc                     &                          &                        \\
Hologram               &                         &                       &                            &                         &                       &                    & \redcirc                         &                        &                       &                    &                          &                          &                    &                           &                          &                        \\
KeepGo                 & \redcirc                     & \redcirc                   &                            &                         &                       & \redcirc                & \redcirc                         & \redcirc                    & \redcirc                   &                    &                          &                          &                    &                           &                          &                        \\
Tago                   &                         &                       & \faDotCircle[regular]                         &                         &                       &                    &                             &                        &                       &                    &                          & \faDotCircle[regular]                       & \faDotCircle[regular]                 &                           &                          &                        \\
Favoriot               & \redcirc                     &                       & \redcirc                        &                         & \redcirc                   &                    &                             & \redcirc                    &                       &                    &                          & \redcirc                      & \redcirc                & \redcirc                       &                          &                        \\
TheThings.io           & \redcirc                     &                       &                            &                         & \redcirc                   &                    &                             & \redcirc                    &                       &                    &                          & \emptycirc                       & \redcirc                &       \emptycirc                    &                          & \redcirc                    \\
Mdash                  & \emptycirc                     &                       & \emptycirc                        & \emptycirc                     & \emptycirc                   &                    &                             & \emptycirc                    &                       &                    & \emptycirc                      & \emptycirc                      & \emptycirc                & \emptycirc                       &                          &                        \\
Fogwing                & \redcirc                     &                       & \redcirc                        &                         & \redcirc                   &                    &                             & \fullcirc                    &                       &                    &                          & \fullcirc                      & \fullcirc                & \fullcirc                       &                          &                        \\
Asksensors             & \redcirc                     &                       & \redcirc                        &                         & \redcirc                   &                    &                             & \redcirc                    &                       &                    & \redcirc                      & \redcirc                      & \redcirc                & \redcirc                       &                          &                        \\
CSL             & \faTimesCircle[regular]                  &                       &                        &                         &            \faTimesCircle[regular]      &               \faTimesCircle[regular]     &                             & \faTimesCircle[regular]                   &                       &                    &                      &                      &                & \faTimesCircle[regular]                       &                          &                        \\
GlobalM2MSIM           & \fullcirc                     &                       & \fullcirc                        &                         & \fullcirc                   & \fullcirc                &                             & \fullcirc                    & \fullcirc                   &                    &                          &                          & \fullcirc                & \fullcirc                       &                          &                        \\
Imvvy                  &                         &                       & \faTimesCircle[regular]                         &                         & \faTimesCircle[regular]                    &                    &                             &                        &                       &                    &                          & \faTimesCircle[regular]                       & \faTimesCircle[regular]                 & \faTimesCircle[regular]                        &                          &                        \\
Open M2M               &                         & \redcirc                   &                            &                         & \emptycirc                   & \emptycirc                &                             &                        & \emptycirc                   &                    &                          &                          &                    & \emptycirc                       &                          &                        \\
ResIOT                 & \fullcirc                     & \fullcirc                   & \fullcirc                        &                         & \redcirc                   & \fullcirc                &                             & \fullcirc                    & \fullcirc                   &                    &                          &                          & \fullcirc                & \fullcirc                       &                          &                        \\
Thingsboard            &                         &                       &                            &                         & \emptycirc                   &                    &                             &                        &                       &                    &                          &                          &                    & \emptycirc                       &                          &                        \\

ResIOT (C)             &                         &                       &                            &                         &                       &                    &                             & \emptycirc                    &                       &                    &                          &                          &                    & \emptycirc                       & \purplecirc                      &                        \\
Bevywise (C)           & \purplecirc                     &                       & \purplecirc                        &                         & \purplecirc                   &                    &                             & \purplecirc                    &                       &                    &                          & \purplecirc                      & \purplecirc                & \purplecirc                       &                          &                        \\

Thingsboard (O)        & \purplecirc                     &                       & \purplecirc                        &                         & \purplecirc                   &                    &                             & \purplecirc                    &                       &                    &                          & \purplecirc                      & \purplecirc                & \purplecirc                       &                          &                        \\
OpenRemote (O)         & \redcirc                     &                       & \redcirc                        &                         & \redcirc                   &                    &                             & \redcirc                    &                       &                    &                          & \redcirc                      & \redcirc                & \redcirc                       & \redcirc                      & \redcirc                    \\
Boodskap (O)           & \redcirc                     &                       & \redcirc                        & \redcirc                     & \redcirc                   &                    &                             & \redcirc                    &                       &                    & \redcirc                      & \redcirc                      & \redcirc                & \redcirc                       &                          &                        \\
DGIOT (O)                  & \purplecirc                     &                       & \purplecirc                        &                         & \purplecirc                   &                    &                             & \purplecirc                    &                       &                    &                          & \purplecirc                      & \purplecirc                & \purplecirc                       &                          &                        \\ \hline
\end{tabular}
\label{tab:shortlistedplatformvulns}
\end{table*}

\subsection{Broken Authentication}
\label{sec:broken-auth}
We found several broken authentication vulnerabilities with varying security consequences: complete account takeover in AskSensors, Fogwing, MDash, Aeris Neo, ResIOT, Favoriot, GlobalM2MSIM and Thingsboard; data exposure in Aeris Neo, Favoriot, TheThings.io and AskSensors; and several IoT SIM registration issues (e.g., denial of service) in Hologram, KeepGo and OneSIMCard. Below we discuss these vulnerabilities, grouped by the affected resources.

\subhead{Account information}
Broken authentication in user account management was found on AskSensors, Fogwing, Boodskap, Aeris Neo, ResIOT, GlobalM2MSIM and Thingsboard. On AskSensors, an attacker can obtain sensitive account information for any user, by providing a 4-digit ID value, %
which can be easily enumerated for all users.
Details such as username, email ID, number of connected IoT devices, account creation date, user's address and the password reset token are exposed. 
Specifically, the reset token can be exploited to completely takeover \emph{any} user account by submitting a forgot password request with the victim's email address.
On Fogwing's analytics portal, an attacker could reset
the password of \textit{any} user by providing the victim's registered email ID. On Boodskap, all APIs return valid responses for unauthenticated requests. Although most of the requests contain UUIDs which cannot be known to a remote attacker, an existing user can abuse this to elevate their role
to an admin. 
An attacker could view sensitive information
of \emph{any} user of Aeris Neo, by providing a 5-digit account ID of the victim. Using a trial and error approach, the attacker can retrieve sensitive information, e.g., name, email ID, account type and API key for other users. The API key leaked in this vulnerability is used for authentication throughout the platform and can be further abused to gather information, e.g., SIM card details, billing details and data usage. Upon submitting a password reset request on GlobalM2MSIM's website, the password reset link
sent to the registered email ID contains a random password reset key. The key parameter is not validated by the platform. Thus, a valid reset link can be made by providing any random string in the key parameter and the email address of the victim user. An attacker could reset the victim user's password and takeover their account, granting access to the IoT SIM cards owned by the victim user. In ResIOT, an attacker could get any user's authentication token
by providing the victim's email address. Thereafter, the attacker could log in to the victim user's account on the platform, and perform IoT SIM management tasks. On MDash, the registration web request
contains a URL parameter; a user-dependent remote attacker can submit a registration request, even for an existing user, and provide an attacker-controlled website in the URL parameter. The platform sends an email to the victim user with the activation link containing the attacker's website URL. Similar attack is possible via password reset requests on Asksensors (by modifying URL parameter in request) and Thingsboard (by modifying Host header in request).

\subhead{IoT device} 
Broken authentication could be abused for remote command execution and device data forgery. An attacker could send arbitrary commands to the IoT devices connected to AskSensors,
where the attacker needs to supply the command and the device ID (a 5-digit numeric value) in a POST request. An attacker could also obtain the GPS coordinates
of each IoT device on the platform. We found broken authentication on Favoriot that could let an attacker obtain sensitive information
such as user ID, Bcrypt hashed password, API keys with read-only and read-write permissions---by providing only the email address of a victim user. %
The leaked API keys could be used to read data sent by the victim's IoT devices, as well as to send forged data on behalf of those IoT devices to the platform. 
On TheThings.io, an attacker could  download the firmware image files
uploaded by \textit{any} user through an unauthenticated web request, by supplying a 5-digit organization-id. 

\subhead{IoT SIM registration}
\label{section:simregistrationprocess}
We found a lack of authentication in the IoT SIM card registration process of 3/6 platforms from whom we purchased IoT SIM cards.
KeepGo users can register their IoT SIM card on the platform by entering their Integrated Circuit Card Identification (ICCID) number; no other verification is required. Upon registration, a request is sent to the platform to check the SIM card's availability.
We refer to such endpoints as IoT SIM validation endpoints. ICCID numbers are susceptible to enumeration as evident from past attacks~\cite{saita}. We altered the last few digits in our own IoT SIM card's ICCID number and observed three types of responses: a) SIM Card Not Found, b) This line already in use, c) Success. We found an unassigned ICCID number within under 100 requests, which we could successfully register from another test account. %
Similarly, 
an attacker can enumerate valid unassigned ICCID (in Hologram)
and SIM numbers (in OneSIMCard)
and
add them to fake accounts. If such an unassigned SIM card
is later purchased by a customer, she would receive an error message during registration of the SIM card (and may require intervention of the support team). 

On Telnyx, registering a SIM card requires the user to provide a 10-digit unique code printed on the physical card. %
In both Telenor and OpenM2M, IoT SIM registration is not performed by the user. Thus, these three platforms are not affected by this issue.

\subsection{Unauthorized Access}
\label{sec:unauthorized-access}
We found that 9/52 platforms are vulnerable to unauthorized access. We provide the details below. %

\subhead{IoT SIM}
On KeepGo and OneSIMCard, an attacker can disrupt IoT SIM cards' cellular connectivity. On KeepGo, users can create sub accounts within their own account and assign IoT SIM cards to these sub accounts. %
An attacker can view other users' sub accounts,
by inputting a 4-digit account ID  (easily enumerated). More importantly, an attacker can deactivate another user's sub account,
which deactivates all the IoT SIM cards assigned to that sub account; the attacker needs to send the victim sub account's ID.
We also found another instance of unauthorized access vulnerability in KeepGo, \textcolor{black}{with} which an attacker can set any arbitrary future date as the date of IoT SIM card deactivation (requires the ICCID number of a targeted SIM, or can be guessed in a trial-and-error approach for arbitrary targets). Both of these attacks can be performed by anyone with a registered account on the platform, with or without an IoT SIM card from KeepGo.

On OneSIMCard, we found four unauthorized access vulnerabilities affecting IoT SIM cards (the first three issues have been fixed). An attacker can block the cellular connectivity
and internet access
for \emph{any} IoT SIM in the platform, by sending web requests with the SIM number of the victim, which is a unique 15-digit numeric value.\footnote{As from the purchased cards, OneSIMCard apparently uses sequential SIM numbers.}
An attacker could remove the IMEI lock
set on any IoT SIM card as well as set arbitrary IMEI lock
on any IoT SIM card by providing the victim's 15-digit SIM number.
An attacker can also set country restrictions for any IoT SIM card,
which would restrict the data access of the IoT SIM card in the countries specified by the attacker. 
Only existing users with SIM cards assigned to their accounts can perform these attacks. 
An attacker can use the broken authentication in SIM registration process (see Sec.~\ref{section:simregistrationprocess})
to obtain such an account.

\subhead{Alerts}
We found unauthorized access vulnerabilities in the alerts configuration on KeepGo and TheThings.io, using which attackers can \textcolor{black}{exfiltrate sensitive information} as well as modify the configured alerts. On KeepGo, users can set rules to trigger an email alert notification if the account balance falls under a set threshold. An attacker could modify this rule
for any user by providing account ID (a 4-digit number), condition value (containing the threshold amount in USD) and the desired email address in a POST request. The attacker could set a large negative value as the threshold and as a result, the alert would not be triggered. An attacker could also redirect the alerts to an arbitrary email address. In both cases, the user would not receive the notification in time, which may lead to service disruption. 
KeepGo users can set rules to
keep track of data usage and detect changes in IMEI corresponding to IoT SIM cards. An attacker can view the rules
set by other users by providing the 3-digit line rule ID. However, the platform blocks unauthorized attempts to modify or delete these rules.

TheThings.io contains a module named \textit{Cloud Code}, allowing users to write jobs, functions, and triggers on the platform~\cite{thethings}. An attacker can view the cloud codes
of other users,  by providing a 5-digit organization ID (easily enumerated). 
TheThings.io lets users utilize third party services e.g., Twilio (for SMS/voice alerts), and SendGrid, Mandrill, SES, and Gmail (for emails) while defining triggers. The unauthorized access vulnerability in \textit{Cloud Code} exposes the authentication credentials (e.g., API keys, tokens) for these third party services as well. %
These credentials can be abused in several ways, e.g., to retrieve sensitive information such as metadata of emails previously sent, to use the service APIs for free (incurred cost will be billed to the victim), and to launch phishing attacks via emails and SMS messages.
Another access vulnerability in the \textit{Alerts Manager} module (which generates alerts depending on user-set conditions), allows an attacker to delete the alerts  set by any user,
leading to potential service disruptions.

\subhead{SMS}
We found unauthorized SMS access vulnerabilities on KeepGo and OneSIMCard platforms, enabling an attacker to view SMS messages exchanged between the IoT devices and the platform, and send arbitrary messages to the IoT devices. KeepGo users can send SMS messages to their IoT SIM cards from the platform, and the responses from the IoT devices can be viewed on KeepGo's website. For each inbound and outbound message, the user is charged 0.05 USD.
An attacker could view all the exchanged IoT SMS messages
by providing the ICCID of the victim's IoT SIM card (see Sec.~\ref{sec:broken-auth}, under ``IoT SIM Registration''), and the billing cycle.
More importantly, an attacker can send arbitrary messages
by inputting the target SIM's ICCID and the SMS text in a POST request, after which, the victim's account balance is reduced by 0.05 USD. Similarly, on OneSIMCard, an attacker could view messages sent to any SIM card, 
and send arbitrary messages to an IoT SIM card by inputting the target SIM number and the SMS text in a POST request, for which the victim's account balance is reduced by 0.01 USD. 
Note that the attacker's account is not charged at all in these attacks. 
On OpenM2M, an attacker could read SMS messages for any SIM card 
by providing a 6-digit subscription ID.

\subhead{IoT device data} %
We found unauthorized access vulnerabilities pertaining to IoT device data with varying implications: arbitrary dashboard modification on TheThings.io; IoT device data forgery in Fogwing and AskSensors; sensitive device data exposure in TheThings.io, Fogwing, AskSensors, and Platform X. 

On TheThings.io, each organization has a dashboard (containing smaller units known as widgets) with an overview of the status of connected IoT devices. An attacker could view the dashboard configuration for any organization
by supplying the victim organization's ID (5-digit numeric value, easily enumerated). The exposed information includes: dashboard ID and configuration of individual widgets on the dashboard (e.g., name of widgets used, source of data). %
More importantly, by using the exposed dashboard ID, an attacker could edit/delete the dashboard configuration
of any organization by inputting the dashboard ID and the desired dashboard settings in a PUT request. 
One of the widgets, \emph{iFrame Link}, lets users load content from external pages in an iFrame. Through unauthorized access, an attacker could add a malicious \emph{iFrame Link} to any organization's dashboard, which could load content from an attacker-controlled website. 
On Fogwing, an attacker could view sensitive information
of any IoT device by providing the 4-digit gateway ID; exposed details include name, edge ID, geolocation, health status, and data received from the device. 
An attacker could further abuse the leaked edge ID to send forged data
on behalf of the targeted IoT device to the platform by providing the victim's edge ID and the attacker's API key as URL parameters along with the forged data payload in the POST request. 

On AskSensors, for any device, simply by providing a 4-digit ID, an attacker could view the API keys,
and remove the device.
An attacker could also read data submitted by any IoT device, and send forged data on behalf of any device to the platform. The exposed API key could be further abused to launch XSS attacks (see Sec.~\ref{section:inputvalidation}).

On Platform X, an attacker could view details of the security updates
deployed on any device by providing the 5-digit update ID in a POST request. Exposed details include device ID, device name, device state (online/offline), update status, update message and update time.
For any device, the last fetched log file could be obtained
by providing the 6-digit device ID in a POST request. An attacker could view outputs of commands issued by other users
(fixed during our testing) by providing their own user token, the command ID and device ID (both 5-digit numbers) 
and their own CSRF token.

\subhead{Account information} We found access control violations leading to exposure of account information on the TheThings.io, Fogwing, ResIOT, AskSensors and Aeris Neo platforms. AskSensors uses Stripe APIs\footnote{https://stripe.com/en-ca} for payment processing, and exposes the stripe card objects\footnote{\url{https://stripe.com/docs/api/cards/object}} of all users, through an API,
where an attacker needs to supply only a 4-digit ID value. Exposed details include card company's name, card expiry month and year, last 4 digits of the card, 
CVC check status (pass, fail, unavailable, unchecked), card type (debit, credit, prepaid) and billing address. 
On TheThings.io, an attacker can obtain organization details
such as organization name and subscription ID for all organizations on the platform. An attacker could also obtain user information
such as the email address and the permissions granted for all users within an organization. In both cases, the attacker needs to provide a 5-digit numeric organization ID value. On Fogwing, an attacker could access details of all the connected enterprises
by providing a 4-digit numeric ID, which can be enumerated easily. Exposed details include enterprise name, business location and type of business. On Aeris Neo, there are two user roles, namely standard user and account manager. Only an account manager can access account management functionality of a given account. 
However, 
all users in an account use the same API key for authentication/authorization. %
A standard user can use the shared API key  
to make themselves the  account manager. 
ResIOT provides a graphical representation of the available credit amount in the user's account. 
An attacker can obtain information about the credit amount
available in \textit{any} account in the past. The request contains a 4-digit numeric id in the URL. 
However, the attack has limited impact since the exact user for a given exposed graph cannot be determined.

\subsection{Vulnerable Trigger-Action Functions}
\label{section:triggerfindings}
In TheThings.io, users can execute JavaScript code in the Jobs, Triggers, and Functions of the \emph{Cloud Code} module on the platform's website. 
These code segments are executed in a sandbox as mentioned in the official documentation~\cite{thethings}. We discovered a remote code execution vulnerability on the platform's server, by following the steps below. Note that the vulnerability affects all the three functionalities (Jobs, Triggers and Functions).

When we used the require module (from Node.js) in a JavaScript function, we obtained an error indicating that its use is disallowed. Therefore, we tried to find more about the sandbox library used by the platform; see %
Sec.~\ref{vuln_trig}. From the stack trace returned in the error message, 
we found that
Jailed~\cite{asvd} sandbox is used. We followed a coordinated disclosure approach, wherein we were provided a perpetual access account by the platform's CEO for further testing. We leveraged known sandbox bypass 
attacks~\cite{asvdissue} against Jailed,
to use require module inside a function to invoke the 
child\_process (from Node.js) module, and test for remote code execution using \texttt{process.exec()}. 
We executed the system command \texttt{id} to confirm that we attained remote code execution with root user privileges. We found that the \textit{Cloud Codes} functionality was running inside a kubernetes pod, shared by all users on the platform. We could attain root access on this shared pod. No further commands were executed which could alter any existing system configuration or \textcolor{black}{exfiltrate any sensitive information.} %

On OpenRemote, users can write rules in Groovy programming language to create event-based workflows. We found that the functionality has not been sandboxed, and allows users to run commands on the server itself. We verified the issue on a local installation on Linux. We were assigned a CVE (rated 9.8/10, critical severity) for this vulnerability. Post disclosure, the functionality has been restricted to super users, and the developers are working on a sandbox implementation.

\subsection{Lack of Input Validation}
\label{section:inputvalidation}
We found SQL injection on ResIOT's locally-deployable platform. We found cross-site scripting (XSS) vulnerabilities on 16/52 evaluated platforms. On 9 of them (OpenM2M, OneSIMCard, Favoriot, TheThings.io, AskSensors, Thingsboard, Boodskap, Bevywise, DGIOT), an attacker could steal session cookies and authentication tokens 
stored in browser's LocalStorage/SessionStorage
from active user sessions via XSS. %
Note that on each of these 9 platforms, we attained stored XSS, providing an attacker perpetual access to session cookies/tokens whenever the user visits the affected pages, leading to account takeover. 
On ResIOT's locally-deployable platform, a user-dependent remote attacker could use cross-site request forgery to execute
arbitrary SQL queries on the platform's database.

On Platform X, an IoT device's log file content is displayed without proper sanitization. Hence, any JavaScript payload injected into the log file can be used to trigger XSS attacks against the corresponding user. An attacker can insert such payloads in devices that may collect attacker-controlled inputs.
On RemoteIOT, users can execute commands to fetch files from the devices. Similar to Platform X, an attacker could insert XSS payloads on devices that collect attacker-controlled inputs (remediated after our  disclosure).

On KeepGo, by exploiting the unauthorized access vulnerability in the SMS functionality (see Sec.~\ref{sec:unauthorized-access}),
an attacker can send a JavaScript payload as an SMS message, which executes when the IoT SIM owner views the list of sent SMS messages on the platform's web portal.
Similarly, an unauthorized access vulnerability in AskSensors (see Sec.~\ref{sec:unauthorized-access}) allows adding JavaScript payload in an IoT device's description field; the payload is executed on the browser when the victim user views the device details. The victim's authorization token stored in the browser's LocalStorage can be stolen via XSS.

On Thingspace's Freeboard portal, a user-dependent remote attacker can abuse CSRF to embed XSS payloads on a user's dashboard. Similar abuse scenarios involving use of CSRF for injecting XSS payloads were found on OneSIMCard and OpenM2M. 
On Thingsboard, an existing user can inject XSS payloads into the user logs, which can also be accessed by an admin user. Thus, any user can steal the admin's authentication token. An existing user can inject XSS payloads on Boodskap, Bevywise and DGIOT to steal admin cookies/tokens.

Broken authentication in Aeris Neo (see Sec.~\ref{sec:broken-auth}) can be used to launch XSS attacks against the users by injecting JavaScript payloads in the first/last name parameters. 
An attacker could abuse broken authentication (see Sec.~\ref{sec:broken-auth})
in Favoriot to make forged data submissions containing XSS payloads.
On TheThings.io and Imvvy, an on-path attacker could capture the authentication credentials from HTTP requests and MQTT(without TLS) messages, respectively and abuse them to launch XSS attacks. 

\subsection{Other Vulnerabilities}
\textcolor{black}{For the other vulnerabilities listed in Sec.~\ref{sec:othervulnsdesc}, here we summarize the corresponding findings (detailed in Appendix~\ref{section:othervulns}). We found cross-site request forgery on 6 platforms, leading to account takeover and partial account modification on 3 platforms each. ResIOT's web-hosted platform was available on both HTTP and HTTPS (HTTP access removed post our disclosure). Five platforms support API calls on HTTP without TLS, enabling API key theft by on-path attackers. Eight platforms are vulnerable to SSLStrip attacks, enabling account takeover by on-path attackers. Secure and HTTPOnly attributes are not set for session cookies on 7 and 10 platforms, respectively. Five platforms return verbose error messages that contain details about platforms' technology stacks and internal directory structures. Five platforms send sensitive information such as IMEI, API keys and ICCID to third parties.}

\section{Discussion}
\label{section:actdiscussion}
In this section, we discuss our limitations, key takeaways from our analysis, and recommendations for platform developers and users. 

\subsection{Limitations}
For 14/20 platforms offering connectivity management services, and for all the platforms offering device and/or data management services, we performed the assessment with trial accounts with a limited set of functionalities. 
Thus our findings may represent a lower bound of the vulnerabilities. 
Several platforms do not allow self-registration of user accounts, and require manual verification (e.g., proof of business ownership) before granting access to the platform; we excluded such platforms. %
Also, most of the vulnerabilities tested and their detection techniques are no different from those adopted in traditional web security. However, results from our vulnerability assessment demonstrate the practical consequenses of such known issues in the IoT management platforms.

\subsection{Key Takeaways}
We draw the following key takeaways and observations from our findings. 

\subhead{Platform-wide attacks} Compared to attacking a single IoT device, the IoT platform provides a large attack surface that encompasses a large number of registered enterprises, their connected devices, and their users~\cite{he2021fingerprinting}. Thus, the impact of vulnerabilities detected in the platforms is amplified significantly. 17/52 platforms are affected by vulnerabilities leading to platform-wide attacks.

\subhead{Platform design flaws} On 9/52 platforms, user resources are not properly isolated, which lead to unauthorized access vulnerabilities. There needs to be a strong binding between the device/SIM IDs and the corresponding users, along with adequate isolation of users and their resources. Also, on 9/52 platforms, the use of short numeric/guessable IDs (e.g., 4-digit values, SIM numbers) instead of longer IDs, coupled with a lack of sufficient rate limiting of web requests, makes it easier for an attacker to quickly launch enumeration/guessing attacks against all IoT entities—enterprises, users, devices. Furthermore, on 3/52 platforms, the absence of unique tokens for registering SIM cards makes it possible for an attacker to register multiple SIM cards which they don’t own. As a result, the customers purchasing those cards would not be able to register and activate them.

\subhead{Multifaceted nature of attacks} Platform issues (broken-authentication/unauthorized-access) can compromise IoT devices, and forged data submissions containing XSS payloads can be used to launch user-specific attacks. This highlights the need for a more carefully considered adversary model. For example, in Asksensors, we found a broken authentication vulnerability on the platform that allows a remote attacker to issue arbitrary commands to any connected device. On the other hand, on TheThings.io, we found that it is possible for on-path attackers to steal device authentication tokens, and further abuse them to launch XSS attacks against the corresponding device owners on the platform.

In addition, we observed that chaining of multiple vulnerabilities can be used to launch attacks with more severe impact. For example, the SQL injection vulnerability on ResIOT's locally-deployable platform could only be performed by platform users. However, by leveraging CSRF, a user-dependent remote attacker could also exploit the SQL injection vulnerability. Additionally, CSRF can be used as a vector to inject XSS payloads on the victim's web dashboard, leading to account takeover (e.g., Bevywise, OneSIMCard and OpenM2M). Similarly, API key theft from devices using insecure communication (HTTP without TLS) can be used to send forged data with XSS payloads, leading to account takeover (e.g., TheThings.io).

\subhead{Unique challenges in locally-deployable platforms} Compared to a Software-as-a-Service (SasS) based web-hosted IoT platforms, locally-deployable platforms are configured and maintained by the enterprises using the platforms. Thus, the administrators are responsible for performing several crucial tasks to ensure security of their deployments. Administrators are required to monitor available patches for newly discovered vulnerabilities and deploy them in a timely manner. In addition, administrators must harden the default installations, which includes changing the default credentials (e.g., passwords, secret keys). Weak, guessable, or hardcoded passwords are the among the top 10 security risks for IoT systems identified by OWASP~\cite{owasp}. Similar to IoT devices, some locally-deployable platforms (e.g., OpenRemote, DGIOT, dojot) have a default user account (typically with admin privileges) that is created upon installation, with a default username and password. If the login credentials for the default account are not changed, an attacker can abuse this configuration flaw to log in to the platform. Furthermore, in case a JSON web token (JWT) is used for platform-wide authentication of an open-source platform (e.g., in Thingsboard, Mainflux), the default value of the signing key used to generate JWT can be obtained from the configuration files. Enterprises deploying any such platform must change the key's value to a non-guessable one; otherwise, an attacker would be able to generate valid JWTs using the default key value (and a target user's email/user ID), and impersonate any user on these platform deployments. Due to ethical considerations, we did not perform an active measurement to determine platform deployments using default credentials.
\subsection{Recommendations} 
We propose the following recommendations for the platform developers and their users. 
(1) Some platforms continue to support devices using cleartext protocols (HTTP, MQTT without TLS). The abuse scenarios 
\textcolor{black}{(see Sec.~\ref{section:inputvalidation})} 
in such cases are deployment-specific (i.e., depends on their users). However, the platforms must emphasize the security implications of such choices, and perhaps adopt strict platform-wide TLS enforcement. 
(2) Data sent by devices using APIs and via MQTT should not be assumed to be secure, and should be appropriately validated and escaped while processing them at the platform's end. 
(3) Custom code functionalities on IoT platforms must run inside sandboxes, which should be secured against sandbox escape attacks. As a defense in depth measure to limit the exploit's impact, the custom functionality should not run with root privileges on the server \textcolor{black}{(see Sec.~\ref{section:triggerfindings})}.
(4) With regards to broken authentication in IoT SIM registration~\textcolor{black}{(as discussed in Sec.~\ref{section:simregistrationprocess})}, the IoT connectivity providers may give a unique and randomized (alphanumeric) token to the SIM card owner when the SIM card is purchased to ensure that only legitimate IoT SIM card owners can add the SIM cards to their management platform accounts. This randomized token could be printed on the SIM card's packaging 
(as observed in case of Telnyx), or sent to the SIM owners via email or SMS. %
(5) The web requests responsible for IoT SIM enumeration must be rate-limited to make it infeasible for an attacker to obtain valid IoT SIM numbers, which have not yet been registered on the platform~\textcolor{black}{(see Sec.~\ref{section:simregistrationprocess})}. 
(6) We observe that only 12/52 platforms have publicly disclosed vulnerability reporting programs such as responsible disclosure and bug bounty; all platforms should adopt such programs. 
(7) Besides fixing the vulnerabilities that we disclosed, platform administrators should also regularly check for new issues, e.g., using our test framework, especially when major code changes are made or when new functionalities are added to the platforms. 
(8) Business/organization users of these platforms may also use our framework for choosing a platform with adequate security support, and for performing periodical security auditing. Our findings can be used to raise awareness of potential security and privacy issues among the users that rely on management platforms for their IoT operations. 

\section{Related Work}
Here, we summarize related past research in IoT cellular connectivity, and IoT security in general.

\subhead{IoT cellular connectivity}
Trend Micro, in collaboration with Europol~\cite{gibson}, studied how IoT SIM cards from compromised IoT devices are misused for committing cyber telecom frauds such as subscription fraud (i.e., abusing business processes to access sensitive data from victim's account), and toll fraud (i.e., initiating high volumes of expensive international calls).
Some cellular connectivity providers offer lower data charges for IoT SIMs compared to non-IoT SIM cards. Past research~\cite{xie2020can} has revealed that it is possible to use IoT SIM cards of such providers outside the IoT devices (e.g., in a smartphone), causing financial loss to the  connectivity providers. Another study~\cite{wang2021insecurity} uncovered several vulnerabilities that can be exploited to launch data and text spamming attacks against IoT SIM card owners. 
Big-data based algorithms have been proposed to detect anomalous behavior in IoT SIM card usage~\cite{zhang2019comprehensive}. 
In a recent BlackHat presentation~\cite{blackhat}, a study on 9 IoT platforms found vulnerabilities such as unauthorized access, insecure communication, and XSS. The work covered the platforms that grant access to the website/APIs only after purchasing SIM cards. Our concurrent study of 20 platforms providing connectivity management services uncovered several other critical attack scenarios (e.g., account takeover, user information disclosure, SIM registration failure, IMEI lock reset) by exploiting vulnerabilities in key platform functionalities. We also found that even with the limited set of functionalities offered to trial accounts with no SIM cards, attackers can launch platform-wide (e.g., obtaining sensitive information of all customers), and user-specific (e.g., account takeover) attacks.

\subhead{IoT security} 
Past research mostly focused on specific domains---e.g., home-automation~\cite{zhou2019discovering, alrawi2019sok, fernandes2016security}, video-surveillance~\cite{obermaier2016analyzing}, smart-buildings~\cite{rondon2021lightningstrike, rondon2022survey}. They relied on vendor-specific devices and mobile companion applications, and as such, did not cover all the platform APIs comprehensively. Our work is device/app-agnostic, and covers both client-side and management APIs---the latter APIs are not covered in prior work. We also cover websites and stand-alone APIs from a variety of IoT platforms: consumer IoT (e.g., Tuya), enterprise IoT (e.g., thethings.io, Kaa IoT), and industrial IoT (e.g., \textcolor{black}{Siemens' Mindsphere}, Fogwing), and more generic IoT platforms (e.g., AWS, Azure). Also, we tested both web-hosted as well as locally-deployable platforms (the latter ones were not analyzed for security vulnerabilities in the past). Past work on trigger-action platforms~\cite{ahmadpanah2021} motivated us to check for sandbox escape issues in trigger-action functionalities on IoT platforms. 
Insecure ecosystem interfaces are included in OWASP's list of top 10 security issues in IoT systems~\cite{owasp}. The websites and web APIs of IoT management platforms are among the most vital interfaces and vulnerabilities in them can be abused to target a large set of users and their devices. IoT platforms have been compared based on the security features mentioned by the platforms in their documentation~\cite{fortino2022iot}; however, no actual security evaluation was performed. In~\cite{singh2015twenty}, the authors discuss twenty security considerations in cloud-supported IoT and the state of research for them. Our study presents an experimental evaluation of vulnerabilities in real-world IoT platforms and how they can be exploited to cause widespread impact. 

\section{Conclusion}
\label{section:conclusion}
We provide a security evaluation framework for IoT management platforms which offer data management, device management and connectivity management (cellular) services for consumer/business/industrial IoT devices. We use our framework to perform a systematic review of 52 real-world IoT management platforms. Our security analysis revealed major unauthorized access flaws in 9 IoT management platforms. We also uncovered other severe vulnerabilities such as broken authentication in 13 platforms, and remote code execution on 2 platforms. 
All the vulnerabilities in our framework can be exploited by attackers with no or minimal expenditure (e.g., to purchase trial SIM cards for connectivity management services).
We hope that our study would help the IoT management platform developers secure their platforms against these easy-to-launch but severe attacks.

\bibliographystyle{IEEEtran}  
\bibliography{mybibliography}

\appendices

\begin{table*}[!htb]
\caption{Overview of the CVE-IDs assigned for vulnerabilities discovered in locally-deployable platforms}
\centering
\begin{tabular}{llrll}
\hline
\textbf{CVE-ID} & \textbf{\begin{tabular}[c]{@{}l@{}}Platform \\ Name\end{tabular}} & \multicolumn{1}{l}{\textbf{\begin{tabular}[c]{@{}l@{}}Base \\ Score\end{tabular}}} & \textbf{Severity} & \textbf{\begin{tabular}[c]{@{}l@{}}Underlying \\ Vulnerability\end{tabular}} \\ \hline
CVE-2022-31860  & OpenRemote                                                        & 9.8                                                                                & Critical          & Missing sandbox for trigger-action function                                  \\
CVE-2022-34020  & ResIOT                                                            & 8.8                                                                                & High              & Cross-site request forgery                                                   \\
CVE-2022-35135  & Boodskap                                                          & 8.8                                                                                & High              & Broken authentication                                                        \\
CVE-2022-34022  & ResIOT                                                            & 7.2                                                                                & High              & SQL injection                                                                \\
CVE-2022-35136  & Boodskap                                                          & 6.5                                                                                & Medium            & Broken authentication                                                        \\
CVE-2022-31861  & Thingsboard                                                       & 5.4                                                                                & Medium            & Cross-site scripting                                                         \\
CVE-2022-35137  & DGIOT                                                             & 5.4                                                                                & Medium            & Cross-site scripting                                                         \\
CVE-2022-34021  & ResIOT                                                            & 5.4                                                                                & Medium            & Cross-site scripting                                                         \\
CVE-2022-35612  & Bevywise                                                          & 5.4                                                                                & Medium            & Cross-site scripting                                                         \\
CVE-2022-35134  & Boodskap                                                          & 5.4                                                                                & Medium            & Cross-site scripting                                                         \\
CVE-2022-35611  & Bevywise                                                          & 4.3                                                                                & Medium            & Cross-site request forgery                                                   \\ \hline
\end{tabular}
\label{table:cves}
\end{table*}

\section{Supplementary Information}
\label{supp-info}
Table~\ref{table:cves} contains the list of CVE-IDs assigned by MITRE for the vulnerabilities we found in locally-deployable platforms.
Table~\ref{table:listoftargets} provides an overview of the analyzed platforms. 
Note that for some services, access to the website/web APIs was restricted to active subscription holders. We did not analyze those services where obtaining such subscription was not feasible.
When the platform seeks company details such as company name and/or company email id while setting up an account, we consider that the account is an enterprise account. Otherwise, we consider that the account is a regular account. The type of account 
relates to the category of users (individual or enterprise) affected by the vulnerabilities found on the platforms.

\section{Other Vulnerabilities}
\label{section:othervulns}
\subsection{Cross-site request forgery}
\label{section:othervulns-CSRF}
Cross-site request forgery (CSRF~\cite{barth2008robust}) 
can lead to user-specific attacks such as account takeover and partial account modification. 
We use CSRF Scanner~\cite{csrfscan} Burp extension to passively detect CSRF vulnerabilities. We remove false positives by checking the presence of custom headers in the web requests.
On OneSIMCard, a user-dependent remote attacker can craft a form for password modification and trick a user to submit the form during their active session, thereby taking over the user account. 
Thereafter, the attacker would be able to perform critical IoT SIM management tasks such as SIM activation/deactivation, IMEI lock removal and IoT SMS delivery. 
On Bevywise and Thingspace's Freeboard portal, CSRF can be abused to inject XSS payload on a user's dashboard; on Bevywise, it leads to account takeover. On ResIOT, an attacker can target the platform's admin via CSRF to perform SQL injection. 
On OpenM2M, an attacker with the knowledge of the victim's 6-digit subscription ID can \textcolor{black}{suspend the victim's IoT SIM card.} However, the attacker would need to include the victim's 6-digit subscription identifier value in the attack form. 
On GlobalM2MSIM, 
\textcolor{black}{an attacker can modify the victim's profile} details such as contact information, delivery address and company VAT number using CSRF. 
Since we did \textcolor{black}{not have an active subscription} for GlobalM2MSIM, we could not validate other abuse scenarios. 

\subsection{Insecure communication}
\label{section:othervulns-insecure}
We check if the platforms use HTTP by default, and if the platforms return valid responses to web requests made over HTTP, or allows MQTT without TLS. 
We also check if a platform is vulnerable to SSLStrip attack by using Bettercap v1.6.2~\cite{bettercap}. 
Upon requesting ResIOT's website using HTTP, the server does not redirect the user to HTTPS (fixed post our disclosure), in which case, sensitive information such as login credentials, SIM details, API tokens are transmitted in cleartext via HTTP. 
Data transmission without TLS is supported on 3 platforms 
via HTTP APIs 
and on 10 platforms 
via MQTT.
If an IoT device is erroneously configured to use HTTP/MQTT (without TLS), the IoT device 
would send data to the platform over cleartext, enabling attacks such as API key theft by on-path attackers.
10 platforms' websites do not set HSTS header in the web responses. 
We use Bettercap to confirm if the SSLStrip attack could be launched on these platforms against the accounts owned by us; 8 platforms are confirmed to be vulnerable (i.e., login username and password can be captured by an on-path attacker). On the remaining 2 platforms lacking HSTS, the attack failed due to the use of custom headers. TLS implementation is deployment-specific for locally-deployable platforms. Using Censys.io, we found\footnote{This is measured by checking standard open ports (port 80/8080 for HTTP and 1883 for MQTT), captured by Censys.io on each deployment as on Sept.\ 1, 2022.} several instances of internet exposed deployments that allow insecure communication via HTTP (without TLS): Thingsboard (1938/2206), DGIOT (119/130), OpenRemote (57/57), Mainflux (26/28), dojot (22/23); several deployments allow insecure communication via MQTT (without TLS): Thingsboard (1446/2206), DGIOT(105/130), OpenRemote (7/57), Dojot (2/23).

\subsection{Misconfigured cookie attributes}
\label{section:othervulns-cookie}
We check the following cookie attributes, which must be set for securing session cookies:
    (a) the \emph{Secure attribute}, which instructs the browser to send the cookie only with HTTPS requests, to prevent cookies from being exposed to on-path attackers; and
    (b) the \emph{HTTPOnly attribute}, to make cookies inaccessible to client-side scripts, for preventing cookies from being stolen via cross-site scripting attacks.
11 web-hosted platforms have inadequately protected session cookies; the Secure attribute, and the HTTPOnly attribute were not set in 7 and 10 platforms, respectively; both attributes were missing in 6 platforms. On locally-deployable platforms, the Secure attribute is linked to TLS implementations (measurement on internet exposed deployments require user logins, not performed due to ethical considerations). HTTPOnly was not set on session cookies of 4 locally-deployable platforms.

\subsection{Information disclosure via error messages}
\label{section:othervulns-error}
Thingspace, ClearBlade, KeepGo, AskSensors and Aeris Neo return verbose error messages upon supplying invalid input in web requests, revealing internal paths and stack trace. 
More importantly, Thingspace leaks the
Base64 encoded authentication credentials (easily decoded to reveal the plaintext username/password) in a verbose error message.
The stack trace indicates that the leaked credentials are used for sending authenticated requests to an internal host (b2bservices.vzwcorp.com). %

\subsection{Information leakage to third parties}
\label{section:othervulns-leakage}
On Hologram, Fullstory (session recording script) captures sensitive information, including credit card details of users (fixed at the time of writing). On Telnyx, customer names and shipping addresses are sent to Fullstory and user details such as name, phone number, company name and address are sent to \url{api-iam.intercom.io}. On \url{TheThings.io}, details sent to \url{api-iam.intercom.io}
include user's name, email, phone number, and the authentication token for the user's devices. 
On Emnify, third-party analytics scripts (Mixpanel and Heapanalytics) capture information such as email ID, organization ID and IMSI.
On SIMcontrol, the API key used for IoT SIM management 
is sent to \url{sentry.io}.
SIMControl responded that they have configured server-side scrubbing of sensitive information on sentry.io.

\begin{table*}[!htb]
\centering
\caption{List of evaluated IoT management platforms. $\checkmark$ indicates that the given service is offered by the platform and analyzed in this work. $\checkmark$* indicates that the service is offered by the platform, but not analyzed in this work; notation used for locally-deployable platforms -- (O): Open-source, (C): Closed-source.}
\begin{tabular}{llllll}
\hline
\textbf{\begin{tabular}[c]{@{}l@{}}Platform \\ Name\end{tabular}} &
  \textbf{\begin{tabular}[c]{@{}l@{}}Platform \\ Type\end{tabular}} &
  \textbf{\begin{tabular}[c]{@{}l@{}}Account \\ Type\end{tabular}} &
  \textbf{\begin{tabular}[c]{@{}l@{}}Connectivity \\ Management\end{tabular}} &
  \textbf{\begin{tabular}[c]{@{}l@{}}Device \\ Management\end{tabular}} &
  \textbf{\begin{tabular}[c]{@{}l@{}}Data \\ Management\end{tabular}} \\ \hline
Azure IoT            & Web-hosted  & Regular &    & $\checkmark$  & $\checkmark$  \\
AWS IoT Core         & Web-hosted  & Regular &    & $\checkmark$  & $\checkmark$  \\
Verizon's Thingspace & Web-hosted  & Enterprise     & $\checkmark$  &    & $\checkmark$* \\
Telus IoT            & Web-hosted  & Enterprise     & $\checkmark$  &    &    \\
Platform X        & Web-hosted  & Regular &    & $\checkmark$  & $\checkmark$  \\
Tuya                 & Web-hosted  & Enterprise     &    & $\checkmark$  & $\checkmark$* \\
Sierra Wireless      & Web-hosted  & Enterprise     & $\checkmark$  & $\checkmark$  & $\checkmark$* \\
Cumulocity           & Web-hosted  & Enterprise     &    & $\checkmark$  & $\checkmark$  \\
Telenor              & Web-hosted  & Enterprise     & $\checkmark$  &    &    \\
Truphone             & Web-hosted  & Enterprise     & $\checkmark$  &    &    \\
Telnyx               & Web-hosted  & Enterprise     & $\checkmark$  &    &    \\
Socketxp             & Web-hosted  & Regular &    & $\checkmark$  &    \\
\textcolor{black}{Siemens' Mindsphere}  & Web-hosted  & Enterprise     &    & $\checkmark$  & $\checkmark$  \\
Aeris Neo            & Web-hosted  & Regular & $\checkmark$  & $\checkmark$* &    \\
Bosch IoT Suite      & Web-hosted  & Enterprise     &    & $\checkmark$  & $\checkmark$  \\
RemoteIOT            & Web-hosted  & Regular &    & $\checkmark$  &    \\
ClearBlade           & Web-hosted  & Enterprise     &    & $\checkmark$  &    \\
OneSIMCard           & Web-hosted  & Enterprise     & $\checkmark$  &    &    \\
Hologram             & Web-hosted  & Enterprise     & $\checkmark$  &    &    \\
Emnify               & Web-hosted  & Enterprise     & $\checkmark$  &    &    \\
Blynk                & Web-hosted  & Regular &    & $\checkmark$  & $\checkmark$  \\
Thinger              & Web-hosted  & Regular &    &    & $\checkmark$  \\
Soracom              & Web-hosted  & Enterprise & $\checkmark$  &    &    \\
KeepGo               & Web-hosted  & Enterprise     & $\checkmark$  &    &    \\
GigSky               & Web-hosted  & Enterprise     & $\checkmark$  &    &    \\
Kaa                  & Web-hosted  & Enterprise     &    & $\checkmark$  & $\checkmark$  \\
Pelion               & Web-hosted  & Regular & $\checkmark$* & $\checkmark$  &    \\
Tago                 & Web-hosted  & Enterprise     &    &    & $\checkmark$  \\
Favoriot             & Web-hosted  & Enterprise &    &    & $\checkmark$  \\
SIMcontrol           & Web-hosted  & Enterprise     & $\checkmark$  &    &    \\
TheThings.io         & Web-hosted  & Enterprise &    & $\checkmark$  & $\checkmark$  \\
MDash                & Web-hosted  & Regular &    & $\checkmark$  & $\checkmark$  \\
Luner                & Web-hosted  & Regular & $\checkmark$  &    &    \\
Fogwing              & Web-hosted  & Enterprise     &    & $\checkmark$* & $\checkmark$  \\
AskSensors           & Web-hosted  & Regular &    & $\checkmark$  & $\checkmark$  \\
CSL                  & Web-hosted  & Regular & $\checkmark$  &    &    \\
GlobalM2MSIM         & Web-hosted  & Enterprise     & $\checkmark$  &    &    \\
Aikaan               & Web-hosted  & Enterprise     &    & $\checkmark$  &    \\
Imvvy                & Web-hosted  & Enterprise     &    &    & $\checkmark$  \\
Open M2M             & Web-hosted  & Enterprise     & $\checkmark$  &    &    \\
ResIOT               & Web-hosted  & Enterprise     & $\checkmark$  &    &    \\
Thingsboard               & Web-hosted  & Enterprise     &    & $\checkmark$   &   $\checkmark$ \\
ResIOT (C)           & Locally-deployable & Enterprise     &    &    & $\checkmark$  \\
Thingsboard (O)      & Locally-deployable & Regular     &    & $\checkmark$  & $\checkmark$  \\
OpenRemote (O)       & Locally-deployable & Regular &    &    & $\checkmark$  \\
Boodskap (O)         & Locally-deployable & Regular &    & $\checkmark$  & $\checkmark$  \\
Bevywise (C)         & Locally-deployable & Regular     &    &    & $\checkmark$  \\
DGIOT (O)            & Locally-deployable & Regular &    &    & $\checkmark$  \\
Mainflux (O)         & Locally-deployable & Regular &    &    & $\checkmark$  \\
Zeus IOT (O)         & Locally-deployable & Regular &    &    & $\checkmark$  \\
IoTGateway (O)      & Locally-deployable & Regular &    &    & $\checkmark$  \\
Dojot (O)            & Locally-deployable & Regular &    & $\checkmark$  & $\checkmark$  \\ \hline
\end{tabular}
\label{table:listoftargets}
\end{table*}

\begin{IEEEbiography}[{\includegraphics[width=1in,height=1.25in,clip,keepaspectratio]{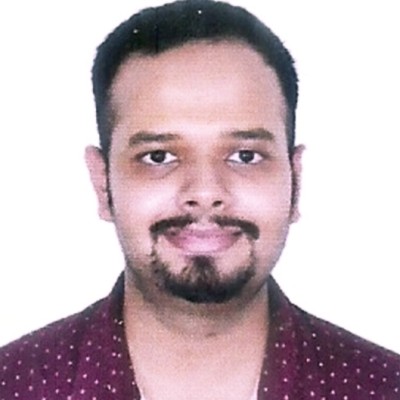}}]{Bhaskar Tejaswi} 

received the M.A.Sc.\ degree in information systems security from Concordia University, Montreal, QC, Canada. In this program, his research focused on two domains, namely, IoT security and phishing. 
He has industry experience in vulnerability assessment and penetration testing, cloud security assessment, and secure configuration review of network devices.

\end{IEEEbiography}

\begin{IEEEbiography}[{\includegraphics[width=1in,height=1.25in,clip,keepaspectratio]{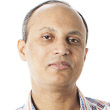}}]{Mohammad Mannan} is a Professor with the Concordia Institute for Information Systems Engineering, Concordia University, Montreal, QC, Canada. His current research interests include Internet and systems security, with a focus on solving high-impact security and privacy problems of today’s Internet.
Dr.\ Mannan is involved with several well-known conferences (e.g., Program Committee: USENIX Security 2022, 2018, ACM CCS 2019, 2016; the Program Co-Chair: ACM SPSM 2016, the General Co-Chair: ACM CCS 2018), and journals (e.g., ACM Transactions on Privacy and Security, the IEEE Transactions on Dependable and Secure Computing, and the IEEE Transactions on Information Forensics and Security).
\end{IEEEbiography} 

\begin{IEEEbiography}[{\includegraphics[width=1in,height=1.25in,clip,keepaspectratio]{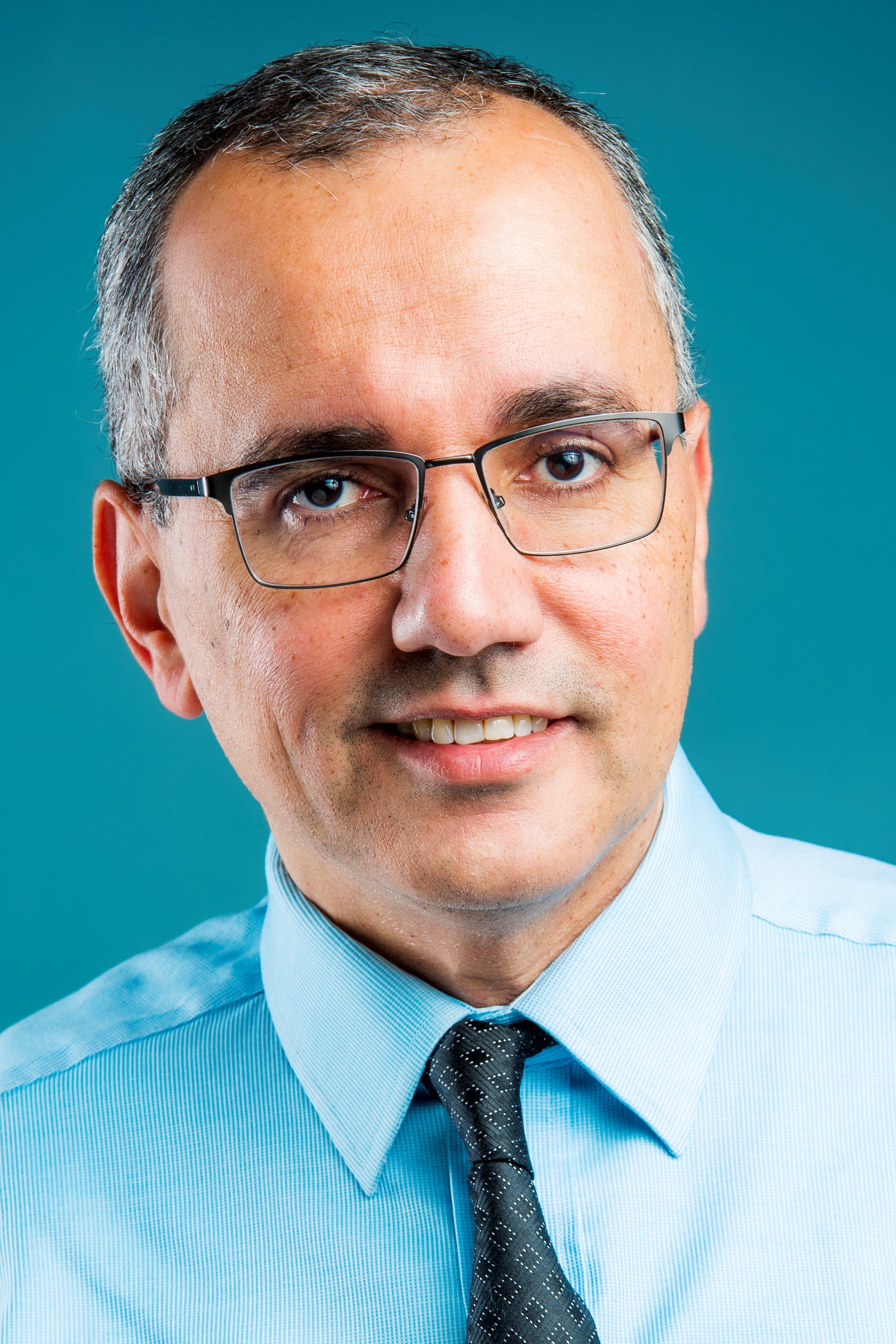}}]{Amr Youssef}
received the B.Sc.\ and M.Sc. degrees from Cairo University, Cairo, Egypt, in 1990 and 1993 respectively, and the Ph.D.\ degree from Queens University, Kingston, ON, Canada, in 1997.  Dr.\ Youssef is currently a professor at the Concordia Institute for Information Systems Engineering (CIISE) at Concordia University, Canada. Before joining CIISE, he worked for Nortel Networks, the Center for Applied Cryptographic Research at the University of Waterloo, IBM, and Cairo University. His research interests include cryptology, cybersecurity, and cyber-physical systems security. 
He has more than 230 referred journal and conference publications in areas related to his research interests.   He was the co/chair for Africacrypt 2013 and Africacrypt 2020, the conference Selected Areas in Cryptography (SAC 2014, SAC 2006 and SAC 2001).
\end{IEEEbiography}

\end{document}